\documentclass[prd,aps,twocolumn,preprintnumbers,amsmath,amssymb,nofootinbib,superscriptaddress,notitlepage]{revtex4-1}

\usepackage{epsfig}
\usepackage[utf8]{inputenc}
\usepackage{color}
\usepackage{epstopdf}
\usepackage[colorlinks=true,
linkcolor=blue,
breaklinks=true,
urlcolor=blue,
citecolor=blue]{hyperref}

\newcommand{\nn}{\nonumber}
\newcommand{\be}{\begin{equation}}
\newcommand{\ee}{\end{equation}}
\newcommand{\bea}{\begin{eqnarray}}
\newcommand{\eea}{\end{eqnarray}}
\newcommand{\ba}{\begin{array}}
\newcommand{\ea}{\end{array}}
\newcommand{\bi}{\begin{itemize}}
\newcommand{\ei}{\end{itemize}}

\newcommand{\lf}{\left}
\newcommand{\rg}{\right}

\newcommand{\ucas}{\affiliation{University of Chinese Academy of Sciences, Beijing 100049, China}}

\newcommand{\imp}{\affiliation{Institute of Modern Physics, Chinese Academy of Sciences, Lanzhou 730000, China}}

\newcommand{\ynu}{\affiliation{Department of Physics, Yunnan University, Kunming 650091, China}}

\newcommand{\giessen}{\affiliation{Institut f\"{u}r Theoretische Physik, Universit\"{a}t Giessen, D-35392 Giessen, Germany}}

\begin{document}

\title{Timelike nucleon electromagnetic form factors: \\  All about interference of isospin amplitudes} 

\author{Xu Cao}\email{caoxu@impcas.ac.cn}
\imp
\ucas

\author{Jian-Ping Dai}\email{corresponding author: daijianping@ynu.edu.cn}
\ynu

\author{Horst Lenske}\email{horst.lenske@physik.uni-giessen.de}
\giessen

\date{\today}

\begin{abstract}
  \rule{0ex}{3ex}
    A striking feature of the timelike nucleon electromagnetic form factors, investigated in $e^+e^- \to N\bar N$ annihilation reactions, is the modulation by local structures of small magnitude and oscillatory form, showing up above $N\bar N$ threshold. Starting from an isospin decomposition of the proton and neutron form factors it is shown that such structures are the natural consequence of the interference of a large and a small amplitudes, resulting in a sinusoidal behavior as a function of the "invariant energy" if the relative phase shift varies with energy. Thus,  periodic oscillations superimposed on a smooth background will be observed.
    In this scenario, an equal size of the modulation for neutron and proton discovered by recent BESIII data evidently implies the {particular} isoscalar or isovector nature of these local structures, or their orthogonal interference, {hence specifies} their origin as excited vector mesons whose widths are tied to the modulation frequency.
    {We clarify that the phase difference {of modulation} between neutron and proton as BESIII data found, but not the modulation itself, is the evidence of an imaginary part of the timelike nucleon electromagnetic form factors, which is associated with the rescattering processes}.
\end{abstract}

\maketitle

\section{Introduction}


A new era of research on nucleon electromagnetic form factors (EMFFs) started with the advent of the modern electron-nucleon and electron-positron facilities.
They allow in an unprecedented manner to investigate fundamental properties of the nucleon in the space- and time-like domains, thus given to access the static and the dynamical properties~\cite{Denig:2012by,Pacetti:2014jai}. Advanced experimental and theoretical methods are allowing investigations of all facets of the tomography of the nucleon - with new prospects for extended studies of the hadron structure at future electron-ion colliders~\cite{Accardi:2012qut,LHeCStudyGroup:2012zhm,Anderle:2021wcy}.
In this paper, we address especially the unsolved issue of the recently identified oscillations seen in the time-like EMFFs~\cite{Bianconi:2015owa,Bianconi:2015vva}.
Various explanations have been proposed~\cite{Bianconi:2015owa,Bianconi:2015vva,Lorenz:2015pba}, but a satisfactory solution of that problem is still pending.
{Especially, the role of vector mesons above $N \bar{N}$ threshold is not fully established though they attracted attention in a microscopic version of the Vector Meson Dominance (VMD) model more than a decade ago \cite{deMelo:2008rj,deMelo:2005cy}.
As a contribution to further understand that phenomenon, we propose methods for investigating the isospin content of these oscillatory structures. }

The timelike EMFFs have been measured in the annihilation process $e^+ e^- \to N \bar{N}$,
whose transition amplitude $G_N$ is complex, connected to the spacelike EMFFs by the dispersion relation~\cite{Belushkin:2006qa}.
The known sources of the imaginary part of amplitudes are multi-meson rescattering loop and intermediate vector meson excitation.
While the former is incorporated by the dispersion analysis of the nucleon EMFF including meson continua~\cite{Belushkin:2006qa,Lin:2021umk,Lin:2021umz,Lin:2021xrc}, the latter is only phenomenologically investigated in the isoscalar and isovector form factor~\cite{Iachello:1972nu}.
{Isospin decomposition is in fact far from trivial when isoscalar and isovector components are both  involved into $e^+ e^- \to N \bar{N}$ reactions.
Two independent isospin channels are insufficient to fully disentangle two complex isospin amplitudes.
Combining available data with isospin decomposition, we show herein that the range of isospin-related parameters could be constrained by available data.
Then we reconcile to a model independent separation of the sources of imaginary part of timelike EMFFs.}

The absolute value of $G_N$ is the so-called nucleon effective form factor (EFF),
\bea \label{eq:eff}
|G_N| &=& \sqrt{\frac{2\tau |G_M(q^2)|^2 + |G_E(q^2)|^2}{2\tau +1}} \,,
\eea
which is explicitly related to the Sachs electric and magnetic form factors $G_{E,M}$. Here $q^2$ is the invariant four-momentum transfer squared, and $\tau = q^2/4m_N^2$ with $m_N$ being the nucleon mass.
The proton EFF
\bea \label{eq:effcomponent}
   |G_p(q^2)| &=&  G_{p}^D(q^2) + G_p^{rsd}(q^2),
\eea
is split into a leading component of modified dipole shape
\bea \label{eq:MODdipole}
G_p^D(q^2) &=& \frac{\mathcal{A}_{p}}{(1+\frac{q^2}{m_a^2})(1-\frac{q^2}{0.71 \textrm{GeV}^2})^2},
\eea
and a residual part $G_p^{rsd}$ accounting for deviations from the smooth behaviour of $G^D_p$. {Since their sum must be a positive real number and by Eq.\eqref{eq:MODdipole} $G^{D}_p$ is a real-valued function, also $G^{rsd}_p$ must be real-valued. $G_n$ is defined accordingly.}

The residual component is chosen of oscillatory shape as a function of the
three-momentum $p =  q \, \sqrt{\tau -1} $ of $N$ (or $\bar{N}$) in the $\bar{N}$ (or $N$) frame~\cite{Bianconi:2015owa}:
\bea\label{eq:osc}
G^{rsd}_N \sim G_N^{osc} &=& A_{N} \textrm{exp}\left(-B_N \, p \right)\, \cos \left(C_N \, p + \it{D_N}\right),
\eea
as imposed by BaBar proton data~\cite{BaBar:2013ukx,BaBar:2005pon} and confirmed also for the neutron by recent high precision measurement of BESIII collaboration between 2.0 and 3.08 GeV~\cite{BESIII:2021dfy}.
Fits to the BaBar data have led to $\mathcal{A}_p = 7.7 \pm 0.3$ and $m^2_a = 14.8$ GeV$^2$~\cite{Bianconi:2015owa,Bianconi:2015vva}.
From the BESIII data, the normalization of the neutron
dipole form factor was obtained as $\mathcal{A}_n = 4.87 \pm 0.09$, using the
same pole mass as for the proton case.
The form chosen for $G_N^{rsd}$ is derived on purely phenomenological grounds.
{The physical origin of the residual component is largely unknown, speculations range from rescattering contributions~\cite{Bianconi:2015owa,Bianconi:2015vva} to yet unknown vector meson resonances and threshold effects~\cite{Lorenz:2015pba}.
Surprisingly the microscopic VMD framework incorporating properly the isospin relation predicted the correct magnitude of the neutron EFF \cite{deMelo:2008rj,deMelo:2005cy}.}
The present status for $G_N^{rsd}$ is shown in Fig.~\ref{fig:Grsd} with $A_p = 0.07 \pm 0.01$, $A_n = 0.08 \pm 0.03$ and sharing of the common parameters for proton and neutron \cite{BESIII:2021dfy}: $B_N = 1.01 \pm0.24 $ GeV$^{-1}$, $C_N = 5.28\pm 0.36$  GeV$^{-1}$.  The phases are $D_p =$ 0.31 $\pm$ 0.17 and $D_n$ = -3.77 $\pm$ 0.55.
It is noteworthy that $D_p$ is consistent with zero within large error bars in a fit to all available $p \bar{p}$ data \cite{Tomasi-Gustafsson:2020vae}.
The parameters of the residual component in oscillatory shape are in fact compatible with constraints imposed by observations and
confirmed by the various other modelling schemes as e.g. in \cite{Tomasi-Gustafsson:2020vae,BESIII:2021dfy}.

\section{Isospin Decomposition and Residual Form Factors}

The striking similarity of the  proton and neutron EFF points to the importance of isospin symmetry -- or more precisely, of charge symmetry -- in the $e^+ e^- \to N \bar{N}$ reactions. Hence, we express the $N \bar{N}$ states  in terms of the  $|I, I_3 \rangle$ = $|0, 0 \rangle$ and $|1, 0 \rangle$ isospin components which allows to represent the proton and neutron form factors with regard to isoscalar ($I_0$) and isovector ($I_1$) amplitudes \cite{Ellis:2001xc}:
\be \label{eq:isoamp}
G_{p,n} = \frac{I_1\pm I_0}{\sqrt{2}} \,,
\ee
where the upper (lower) sign applies to the proton (neutron) case. The above mixtures of isoscalar and isovector amplitudes of equal relative weight but different sign are imposed by isospin symmetry as expressed by the underlying Clebsch--Gordan coefficients. Hence, proton and neutron form factors will be different provided both isospin amplitudes are of comparable size.

The experimentally observed oscillatory structures are taken into account by decomposing the isospin transition amplitudes explicitly into smooth leading components $I^D_{0,1}$  and residual components $I^{rsd}_{0,1}$: 
\bea
 G_p &=& G^D_p + G^{rsd}_p = \frac{I_1^D + I_0^D}{\sqrt{2}}  + \frac{I_1^{rsd}+I_0^{rsd}}{\sqrt{2}} \,,  \\
 G_n &=& G^D_n + G^{rsd}_n = \frac{I_1^D - I_0^D}{\sqrt{2}}  + \frac{I_1^{rsd}-I_0^{rsd}}{\sqrt{2}} \,,
\eea
also expressing the residual amplitudes in an isospin--symmetric ansatz.

In general, the isospin form factors $I^{D,rsd}_{0,1}$ will be complex numbers which is taken onto account by using without loss of generality
\bea \label{eq:ampcplxp}
I_1^D + I_0^D &=& \sqrt{2} |G^D_p| \, e^{i \phi_p^D} \,, \\
I_1^D - I_0^D &=& \sqrt{2} |G^D_n| \, e^{i \phi_n^D} \,, \label{eq:ampcplxn}
\eea
while
\bea \label{eq:ampRSDxp}
I^{rsd}_p&=&I_1^{rsd} + I_0^{rsd} =  |I^{rsd}_p| \, e^{i \phi^{rsd}_p} \,, \\
I^{rsd}_n&=&I_1^{rsd} - I_0^{rsd} = |I^{rsd}_n| \, e^{i \phi_n^{rsd}} \,, \label{eq:ampRSDxn}
\eea
where the phases $\phi_{p,n}^D(q^2)$, $\phi_{p,n}^{rds}(q^2)$ and the amplitudes $|I^{rsd}_{p,n}|(q^2)$ depend on energy and have yet to be determined.
By means of Eq.\eqref{eq:effcomponent}, we obtain
\bea
|G_p|^2 -(G_p^D)^2&=& G_p^{rsd}(2G_p^D+G_p^{rsd})\nn \\
&=& \frac{1}{2} |I^{rsd}_p|^2 +\sqrt{2}|G^D_p||I^{rsd}_p|\Re \lf[ e^{i (\phi_p^D-\phi^{rsd}_p)}  \rg] \nn \,, \label{eq:Gp2}  \\
|G_n|^2 -(G_n^D)^2&=& G_n^{rsd}(2G_n^D+G_n^{rsd})\nn \\
&=& \frac{1}{2} |I^{rsd}_n|^2 +\sqrt{2}|G^D_n||I^{rsd}_n|\Re \lf[ e^{i (\phi_n^D-\phi^{rsd}_n)}  \rg] \nn \,. \label{eq:Gn2}
\eea
Thus, we have found equations relating $G^{rsd}_{p,n}$ to the isospin amplitudes $I^{rsd}_{p,n}$. The leading order leading solutions are
\bea
G_p^{rsd} &\simeq& \frac{|I^{rsd}_p|}{\sqrt{2}}\cos(\phi_p^D-\phi_p^{rsd}) \quad \label{eq:pexpansion} \\
G_n^{rsd} &\simeq& \frac{|I^{rsd}_n|}{\sqrt{2}}\cos(\phi_n^D-\phi_n^{rsd}) \quad \label{eq:nexpansion}
\eea
where terms  of order $|I_{p,n}^{rsd}/G^D_{p,n}|\ll 1$ are neglected. Thus, the form factors $G^{rsd}_{p,n}$ are given in magnitude by the corresponding residual isospin form factors, modulated by a sinusoidal energy dependence introduced by the interference with the leading (smooth) components $G^D_{p,n}$.

A closer inspection reveals, that the form factors $G^{rsd}_{p,n}$ are given in fact by the real parts of the interference term of the corresponding leading and the residual nucleon form factors.
The imaginary part of interference terms is accessible through polarized asymmetries as pointed out in \cite{Tomasi-Gustafsson:2005svz,Haidenbauer:2014kja}. As addressed in the Introduction, the pattern of these interference structures were already highlighted in \cite{Bianconi:2015owa} based on the BaBar measurements~\cite{BaBar:2013ves} and confirmed by the high precision data of the BESIII collaboration~\cite{BESIII:2019hdp,BESIII:2021aer}.

\section{VMD model and Isospin}

By Eqs.~(\ref{eq:pexpansion},\ref{eq:nexpansion}) the EMFF of proton and neutron should display sinusoidal modulations of a similar pattern, although in general differences in modulus and phase have to be expected because  isoscalar and isovector components are superimposed differently. The recent BESIII data disclosed surprisingly that they are of the same magnitude over a wide energy range~\cite{BESIII:2019hdp,BESIII:2021dfy} implying
\be \label{eq:pureiso}
\frac{|I_1^{rsd}+I_0^{rsd}|}{|I_1^{rsd}-I_0^{rsd}|} = \frac{A_{p}}{A_{n}} = 0.88 \pm 0.35 \,,
\ee
where the large error is mainly due to the less accurate neutron data. Its consistency with unity means that either $I^{rsd}_{0} = 0$ or $I^{rsd}_{1} = 0$  or -- as an unlikely third option -- a vanishing interference term by a coincidental phase difference of $\pi/2$ or odd--numbered multiples thereof.

However, a decisive role is played by the production mechanism {hidden in the photon-nucleon-anti-nucleon vertex}. If the $N\bar N$ channel is populated by the decay of an intermediate (off--shell) isoscalar meson as e.g. the $\phi(2170)$ \cite{Zyla:2020zbs}, isospin conservation in strong interaction processes restricts the production process to the $I=0$ component. If an intermediate isovector meson is involved, e.g. $\rho(2150)$  \cite{Zyla:2020zbs,deMelo:2008rj,deMelo:2005cy} of a width comparable to the one of $\phi(2170)$, the $I=1$ components of the $N\bar N$ channels are  populated exclusively. In either of these "isospin-clean" {VMD} scenarios, the states will display distributions of the same spectral shapes in $p\bar p$ and $n\bar n$ channels, except for minor distortion due to isospin symmetry breaking by electromagnetic (and weak) interactions. Hence, the observed similarity of EMFF in $e^+e^-\to p\bar p$ and $e^+e^-\to n\bar n$ reactions leads to the conclusion that the final channel are populated in an isospin--clean production process.

The general case of physical relevance will be given by isoscalar and isovector form factors composed of
superposition of the various isospin-clean {VMD process}, accessible by the available c.m. energy. In particular, the residual amplitudes become
\bea
I_{0,1}^{rsd} &=& \sum_k \delta^k_{0,1} e^{i \phi^k_{0,1}} f^k_{0,1}
\eea
with energy independent relative phases $\phi^k_{0,1}$,  amplitudes $\delta^k_{0,1}$ of magnitudes much lesser than $|I^D_{0,1}|$, and spectral distributions $f^k_{0,1}$.
{ The $\delta^k_{0,1}$ and $f^k_{0,1}$ are the same for $p \bar{p}$ and $n \bar{n}$ channels, which is a direct consequence of above isospin relation.} Within each isospin channel, the overlapping spectral distribution may eventually lead to destructive interference at certain energies and phase--coherent enhancements in other energy regions. However, unfortunately the present data do not yet allow to identify the isospin character underlying the production process.
{Herein a single $f^k$ is used to describe one local structure for simplicity.
This hypothesis does not invade the following discussion and conclusion.}
Specific experiments and theory input are indispensable in order to explore the multi--component spectral structure of the whole set of isospin amplitudes.

\begin{table*}[tb]
    \caption{\label{tab:values}
      {Parameters of three local structures below 2.5 GeV in the fit of Breit-Wigner distribution (or Gauss distribution in the parentheses), together with the extracted $\Gamma_{ee}\Gamma_{N\bar{N}}$ for Breit-Wigner distribution. The $\chi^2/d.o.f$ is 0.7 (1.2). }
    }
    \begin{ruledtabular}
  \begin{tabular}{l c c c}
$k$/BW(GS)        &      1          &       2          &   3              \\ \hline
$M_k$ (MeV)       &    1910$\pm$10 (1958$\pm$10)   &     2083$\pm$27 (2148$\pm$16)  &   2328$\pm$22 (2365$\pm$13)    \\

$\Gamma_k$ (MeV)  &      {32$\pm$32} (30$\pm$8)  &     162$\pm$55 (60$\pm$17)  &   162$\pm$57 (52$\pm$14)    \\

$\delta^k$        &-0.072$\pm$0.048 (-0.036$\pm$0.010) & 0.041$\pm$0.007 (0.024$\pm$0.005) & -0.032$\pm$0.005 (-0.027$\pm$0.007) \\

$\phi_p^D-\phi^{k}$ &   0             & 0.764$\pm$0.373 (0.235$\pm$0.307)   &  1.558$\pm$0.310 (0.046$\pm$0.241)  \\
\hline
$\Gamma_{ee}\Gamma_{N\bar{N}}$ (keV$^2$)  & {4 $\pm$ 4} & 80 $\pm$ 47 & 62 $\pm$ 37
     \end{tabular}
  \end{ruledtabular}
  \end{table*}

\begin{figure}[t]
    \centering
    \includegraphics[width=\linewidth]{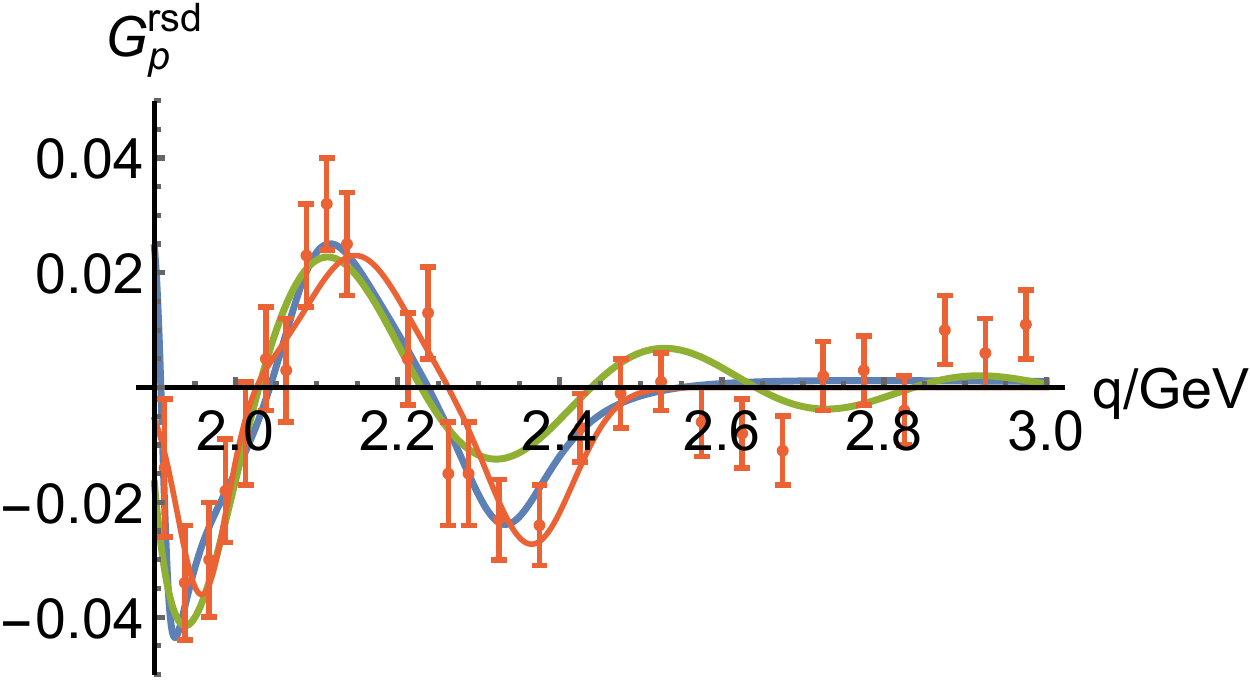}
    \includegraphics[width=\linewidth]{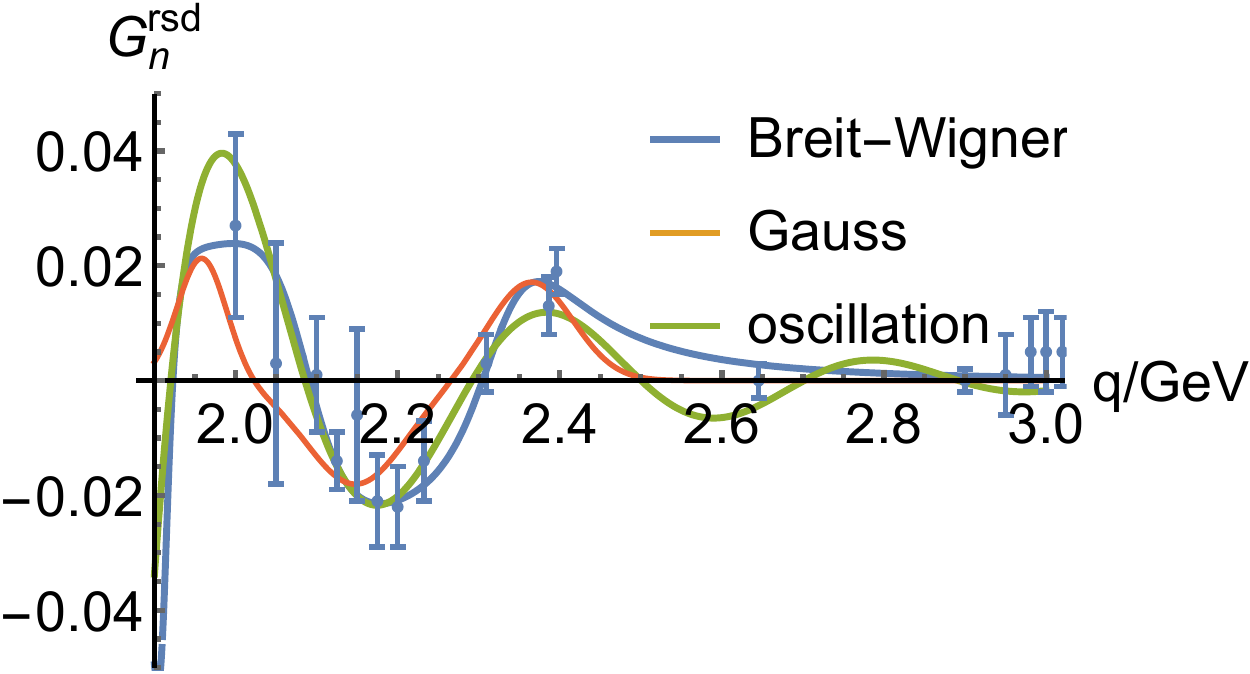}
    \caption{The fit of Breit-Wigner distribution and Gauss distribution to three local structures below 2.5 GeV in comparison with BESIII data~\cite{BESIII:2019hdp,BESIII:2021dfy}.}
    \label{fig:Grsd}
\end{figure}

\section{A Practical Approach to Data Analysis}

A physically well motivated ansatz for a spectral distribution mimicking  the line shape of the residual component is the Breit--Wigner (BW) distribution:
\bea \label{eq:BW}
f^k_{\textrm{BW}} (q^2) &=&  \frac{M_k\Gamma_k}{q^2-M_k^2  + i M_k\Gamma_k}\\
&=&\frac{e^{-i\phi^k_{\textrm{BW}}}M_k\Gamma_k}{|q^2-M_k^2  + i M_k\Gamma_k|},
\eea
where
\be
\phi^k_{\textrm{BW}}= \textrm{arccot} \frac{q^2-M_k^2}{-M_k \Gamma_k}
\ee
which is consistent with the VMD-inspired model with proper analytic continuation~\cite{Iachello:1972nu,deMelo:2008rj,deMelo:2005cy}. Within the VMD interpretation, the local structures are coherent superpositions of vector mesons of isoscalar $\phi/\omega$ or isovector $\rho$ character, located above the $N\bar N$ threshold and with a width $\Gamma_k$ of around 100 MeV, found in \cite{Zyla:2020zbs} and used in \cite{Lorenz:2015pba,Lin:2021xrc}.
Thus, in the BW--approach the residual components carry the energy dependent phases
\be \label{eq:BWphase}
\phi_N^D-\phi_N^{rsd} = \phi_N^D - \phi_N^k  -   \phi^k_{\textrm{BW}},
\ee
where the BW--widths provide the scale for the momentum frequency in Eq.~(\ref{eq:osc}):
\be
C_N \sim \frac{1}{\Gamma_k}.
\ee

Another frequently used alternative, also considered in the fits to data, is to approximate the spectral functions by  Gauss (GS) functions:
\bea\label{eq:GS}
f^k_{\textrm{GS}} (q^2) &=&  e^{-\frac{(\sqrt{q^2}-M_k)^2}{2\,\Gamma_k^2}}
\eea
together with $\phi^k_{p,n}=const.$. With a  proper choice of parameters the Gaussians lead to distributions of shapes closely resembling the Breit-Wigner distributions. In this case, the phase differences $\phi_N^D-\phi_N^{rsd} = \phi_N^D -\phi_N^k$ are constants and can be chosen to be zero. That choice is supported by the fitted values in Tab.~\ref{tab:values}, which are consistent with zero, however, within large error bars.

Application to the BESIII neutron data~\cite{BESIII:2021dfy} reveals that
\bea \label{eq:phasedata}
\phi_n^D- \phi_p^D+ \phi_p^{rsd} -\phi_n^{rsd} = 4.08 \pm 0.58 \, \textrm{rad} \,, \quad
\eea
is a constant deviating from $3\pi/2$ by $ 0.63 \pm 0.58$ rad $= 36.2^\circ \pm 33.2^\circ$,
{where in fact the part $\pi$ of the phase trivially reflects the overall relative sign between $I^{rsd}_p$ and $I^{rsd}_n$.
The phases $\phi_{N}^{rsd}$ of the residual components fully account for the energy dependence as imposed by the data.
The constant value of the total phase difference implies that $\phi^D_{n}-\phi^D_{p}$ behaves
as a function of energy up to a constant complementary to the difference of residual phases.
Assuming that a least one the phases $\phi^D_{p,n}$ is non--zero, we have direct evidence
for the in general complex--valued character of the form factors of Eq.(8) and Eq.(9), respectively. That result is not unexpected, but here probably for the first time confirmed by data. }

It is tempting to interpret the results shown in Fig. \ref{fig:Grsd} in terms of physical processes. Only three clearly developed local structures in the region below $q$ = 2.5 GeV are included in our fit, as shown in Fig.~\ref{fig:Grsd}. The phase difference between neutron and proton is fixed at the central value in Eq.(\ref{eq:phasedata}) in order to reduce the number of free parameters.
Surprisingly,  the parameters in Table~\ref{tab:values} are indeed showing a compatibility with the vector meson spectrum.

The first structure just below 2~GeV  could be a threshold effect due to the opening of the $N\bar{N}$ channel~\cite{Milstein:2018orb}. However, isovector processes involving $\rho(1900)$~\cite{Zyla:2020zbs} are favored in fact by the specified isospin of the structure. The second structure just below 2.2~GeV corresponds energetically to an isoscalar process running through $\phi(2170)$, whose properties are still a matter of intense investigation~\cite{BESIII:2020vtu}, competing with an isovector process driven by $\rho(2150)$~\cite{deMelo:2005cy,deMelo:2008rj}.
The third structure close to 2.4~GeV is possibly produced by $\rho(2350)$~\cite{Zyla:2020zbs} which would indicate the population of the isovector $I=1$ component of the proton and neutron EMFF. The state may also be coupled strongly to $\Lambda \bar{\Lambda}$ as found in $p \bar{p} \to \Lambda \bar{\Lambda}$~\cite{Bugg:2004rj} and $e^+ e^- \to \Lambda \bar{\Lambda}$~\cite{Cao:2018kos}. However, these interpretations are suffering from large uncertainties and must be taken with care, especially waiting for the safe confirmation of the vector meson spectrum above 2.0 GeV~\cite{Wang:2021abg}.

An additional important item characterizing the reactions is the partial width for decay into a specific channel. The product of partial widths is determined by
\bea \label{eq:widthpro}
\Gamma_{ee}\Gamma_{N\bar{N}} &=& \frac{\alpha_{em}^2\beta}{18}\left(1 + \frac{1}{2\tau} \right)  (\delta^k \Gamma_k)^2, \,
\eea
where $\alpha_{em}$ is the electromagnetic fine structure constant, $\beta$ is the center-of-mass velocity of the final state (anti-)nucleon.
The extracted values of $\Gamma_{ee}\Gamma_{N\bar{N}}$ are shown in Table \ref{tab:values} and should be compared to data for $J/\psi$ and $\psi(3686)$ amounting to (1.09 $\pm$ 0.05) keV$^2$ and (2.02 $\pm$ 0.10) keV$^2$, respectively.

A ratio which contains no residual component and Coulomb correction is obtained by
\be
R^D_N = \frac{\sigma^D_n}{\sigma^D_p/C} =|\frac{G^D_n}{G^D_p}|^2  = 0.40 \pm 0.03,
\ee
where $C(q^2)$ is the S-wave Sommerfeld-Gamow Coulomb correction factor \cite{Sakharov:1948plh}, which is equal to 1 for neutron.
Surprisingly, $R^D_N$ is constant over a wide range of energies. This value is close to 3/7, a prediction of SU(6) symmetric nucleon wave function~\cite{Farrar:1975yb}. It lies between the naive $e_u^2/e_d^2 = $1/4 from the quark charge ratio~\cite{Chernyak:1984bm} and 2/3 obtained from the constituent quark model~\cite{Farrar:1975yb}. In any case, the significant deviation from unity is a clear signature of its difference from the residual component, see Eq.~(\ref{eq:pureiso}). 

An isospin relevant ratio is:
\be \label{eq:dataRID}
R_I^D = \frac{|G^D_p|^2 - |G^D_n|^2}{|G^D_p|^2 + |G^D_n|^2} = \frac{2\,\Re (I_0^D I_1^{D\dag})}{|I_0^D|^2 + |I_1^D|^2} = 0.43 \pm 0.03, \quad
\ee
which means that the interference between $I_0$ and $I_1$ is constructive and the isoscalar $I^D_0$ and isovector $I^D_1$ amplitudes are of comparable magnitude.
Hence , we may relate the two isospin amplitudes by
$I_1 = I_0 \delta_I \, e^{i \phi_I} $, rendering Eq. (\ref{eq:dataRID}) to
\be
R_I^D =\frac{2\delta_I \cos{\phi_I}}{1 + \delta_I^2}\leq |\cos{\phi_I}|
\ee
which allows to derive constraints on the allowed ($\delta_I,\phi_I$) combinations. The correlation diagram, obtained by imposing the value for $R^D_I$ derived above is shown in Fig.~\ref{fig:delta-phi}. The results are easily converted to a constraint on $\phi_p^D-\phi_n^D$ introduced in Eqs.(\ref{eq:ampcplxp},\ref{eq:ampcplxn}).

\section{Flavor decomposition}

\begin{figure}[t]
    \centering
    \includegraphics[width=\linewidth]{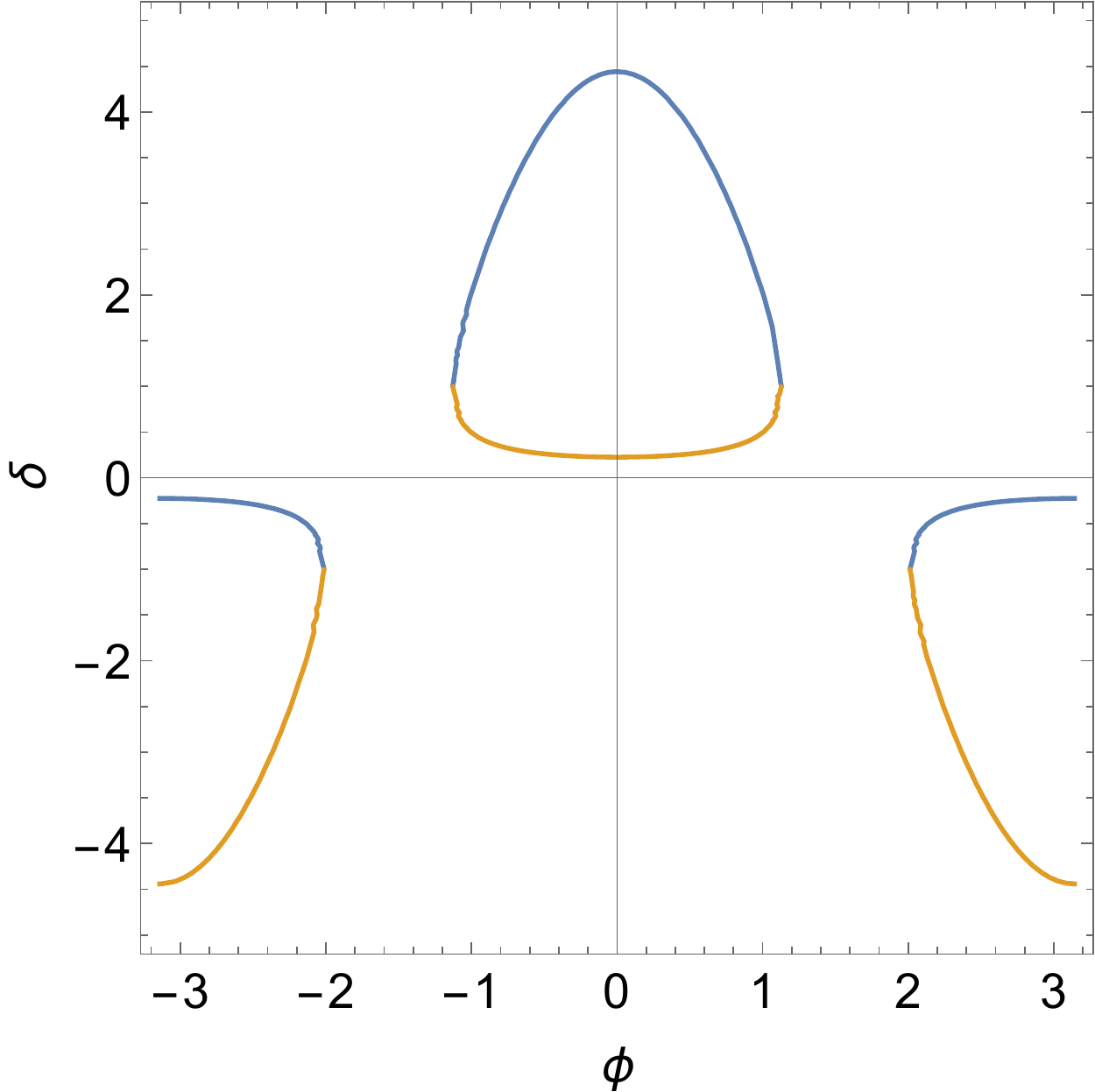}
    \caption{Present data constraints for the relative moduli $\delta = \delta_I$ (or $\delta_M$) and relative phase $\phi = \phi_I$ (or $\phi_M$) of nucleon isoscalar and isovector amplitudes (or form  factors).}
    \label{fig:delta-phi}
\end{figure}

In the isospin limit, when the masses of the up and down quarks are degenerate, the isospin relation is a realization of the flavor decomposition of nucleon EMFF~\cite{Alexandrou:2018sjm},
\bea
G^p_{E,M} &=& \frac{2}{3} G^u_{E,M} - \frac{1}{3} G^d_{E,M} = \frac{1}{2} (\frac{G^{u+d}_{E,M}}{3} + G^{u-d}_{E,M}), \quad \\
G^n_{E,M} &=& \frac{2}{3} G^d_{E,M} - \frac{1}{3} G^u_{E,M} = \frac{1}{2} (\frac{G^{u+d}_{E,M}}{3} - G^{u-d}_{E,M}), \quad
\eea
if isospin (or charge) symmetry $G^{u/n} = G^{d/p}$ and $G^{d/n} = G^{u/p}$ is given. We define the effective isovector ($u+d$) and isoscalar ($u-d$) form factors:
\be
|G^{u\pm d}_{\textrm{eff}}| = \sqrt{\frac{2\tau |G^{u\pm d}_M(q^2)|^2 + |G^{u\pm d}_E(q^2)|^2}{2\tau +1}}.
\ee
Then the sum of the proton and neutron EMFF could be expressed in a compact way,
\bea
|G^D_p|^2 + |G^D_n|^2 &=& \frac{1}{2} \lf( |\frac{G^{u+d}_{\textrm{eff}}}{3}|^2 + |G^{u-d}_{\textrm{eff}}|^2 \rg)
\eea
as a measure of incoherent sum of effective isovector and isoscalar form factors. The non-zero difference between them
\bea
|G^D_p|^2 - |G^D_n|^2 &=&  \frac{1}{3} \Re \lf[ \frac{2\tau G^{u + d}_M G^{u - d\dag}_M + G^{u + d}_E G^{u - d\dag}_E }{2\tau +1} \rg] \qquad
\eea
shows that the interference between these complex form factors is large, arriving at the same conclusion above. For further demonstration we need to assume $G_E = G_M$, so $|\cos\phi_M| \geq R_I^D$ in terms of the notation $G^{u + d}_M = G^{u - d}_M \delta_M \, e^{i \phi_M} $. Then the range of parameters is the same as in Fig.~\ref{fig:delta-phi}. In the SU(3) Nambu--Jona-Lasinio model, the difference between the isoscalar and isovector form factors is shown to be quite small \cite{Kim:1995mr}.

\section{Summary and Perspectives}

{The structure of the nucleon form factors in the timelike region has been investigated in an isospin formalism addressing especially the still puzzling oscillatory contributions, being observed after subtraction of the leading order dipole--type form factors. In our approach, the oscillations are obtained as the natural consequence of the interference of the dipole form factors with a residual component assumed to be of dynamical origin with an intimate relation to the production mechanism in $e^+e^-$--annihilation reactions. In a first application to recent BESIII data we could indeed correlate the observed oscillatory structures to
{the VMD model} proceeding through intermediate virtual vector mesons in the energy region of interest, just above the $N\bar N$--threshold.}

Although neither wide or narrow vector mesons of light quarks are found to couple strongly to the $N \bar{N}$ channels, but they still would be weakly present in timelike nucleon EMFF.
The well known interference effects in unpolarized cross sections were exploited for extracting quantitative results on the  dynamical contributions to the timelike nucleon form factors.
 Since these residual contributions are of much smaller magnitude than the bulk dipole components, the resulting spectral shapes can be expanded order by order into harmonics of the energy dependent phase. The oscillatory behavior observed in the $e^+ e^-$ data below 3.0 GeV is a realization of this phenomenon, described already convincingly well by the leading order harmonics.

{An appealing aspect of the combination of the isospin formalism with the {VMD model} is that it leads to a simple explanation
of the -- up to an overall sign -- almost identical oscillation pattern for protons and neutrons, found in the BESIII data. Within our formalism, the similarity of the periodic structures is a signature of the production process, imprinting the isospin of the virtual meson state into the final nucleon-anti-nucleon configurations which, as a result, are produced selectively via their "isospin-clean" $I=0$ or $I=1$ components.}

Modelling the mesonic spectral functions by a Breit--Wigner distribution, we have shown that the oscillations are in fact a signal of the imaginary parts of the amplitudes. The finite widths -- or likewise the finite lifetimes -- of the vector mesons alone are already sufficient to explain the observed structures. Hence, we conclude that the frequently discussed multi--meson rescattering processes seem to be of minor importance for these structures. However, traces of such interactions are contained in the decay properties and resulting self-energies of the involved heavy mesons. Moreover,
the phase difference between proton and neutron EFF, found by BESIII collaboration, may be a signature of a non-trivial imaginary part of a rescattering amplitude contributing to the smooth dipole component.

Hence, two sources can be identified for an imaginary part in the timelike nucleon EMFF~\cite{Bianconi:2015owa,Lorenz:2015pba} based on available data. We emphasize that after subtraction of the residual component the ratio of the proton to the neutron EFF is surprisingly constant over a wide range of energies, indicating a balanced isospin content since neither the isoscalar or isovector component is dominant. Within error bars, the magnitude of that ratio is consistent with SU(6) predictions.
For future work it is tempting to investigate by the same approach also the timelike EMFF of other octet baryons, where the $\Sigma$~\cite{BESIII:2020uqk,BaBar:2007fsu} and $\Lambda$~\cite{BESIII:2021ccp} hyperons are of special interest.

Last but not least we address the question of higher order expansions involving higher order harmonics $\sim \cos [n(\phi_N^D-\phi_N^{rsd})]$ of rank $n > 1$. However, the currently achievable accuracies of the measurements does not allow safe conclusions on such contributions, mainly because of the smallness of the residual amplitude. For such purposes deep virtual Compton scattering, which interferes with the Bethe--Heitler process~\cite{Belitsky:2001ns} is much more suitable for specific investigations of cross sections in terms of "angular harmonics", where induced higher harmonics have already been observed.

In our approach, the oscillations are induced by the broad vector mesons which, however, are coupled weakly to the $N \bar{N}$  channels. A somewhat more favorable situation is encountered in
the charmonium region. The $J/\psi$ and $\psi(2S)$ states are very narrow and they decay significantly into $p \bar{p}$ and $n \bar{n}$ with equal strength~\cite{ParticleDataGroup:2020ssz}, indicating isospin--clean population in these cases via the $I=0$ components of $N\bar N$ channels. Experimental studies in the vicinity of $\psi(3770)$, where a deep dip is clearly visible in $p \bar{p}$ channel, would be the ideal tool for investigating higher harmonics~\cite{BESIII:2014fwz}, especially if also data for the $n \bar{n}$ channel above $D\bar{D}$ threshold were available with high precision.

\bigskip

\begin{acknowledgments}

Jianping Dai are grateful to Xiaorong Zhou and Dexu Lin for useful discussion and help on experimental measurements. This work is supported by the National Natural Science Foundation of China  (Grants Nos. 12075289, U2032109 and 12165022) and the Strategic Priority Research Program of Chinese Academy of Sciences (Grant NO. XDB34030301).

\end{acknowledgments}


\begin{thebibliography}{1}

\bibitem{Denig:2012by}
A.~Denig and G.~Salme,
Prog. Part. Nucl. Phys. \textbf{68}, 113-157 (2013)
doi:10.1016/j.ppnp.2012.09.005
[arXiv:1210.4689 [hep-ex]].

\bibitem{Pacetti:2014jai}
S.~Pacetti, R.~Baldini Ferroli and E.~Tomasi-Gustafsson,
Phys. Rept. \textbf{550-551}, 1-103 (2015)
doi:10.1016/j.physrep.2014.09.005

\bibitem{Accardi:2012qut}
A.~Accardi, J.~L.~Albacete, M.~Anselmino, N.~Armesto, E.~C.~Aschenauer, A.~Bacchetta, D.~Boer, W.~K.~Brooks, T.~Burton and N.~B.~Chang, \textit{et al.}
Eur. Phys. J. A \textbf{52}, no.9, 268 (2016)
doi:10.1140/epja/i2016-16268-9
[arXiv:1212.1701 [nucl-ex]].

\bibitem{LHeCStudyGroup:2012zhm}
J.~L.~Abelleira Fernandez \textit{et al.} [LHeC Study Group],
J. Phys. G \textbf{39}, 075001 (2012)
doi:10.1088/0954-3899/39/7/075001
[arXiv:1206.2913 [physics.acc-ph]].

\bibitem{Anderle:2021wcy}
D.~P.~Anderle, V.~Bertone, X.~Cao, L.~Chang, N.~Chang, G.~Chen, X.~Chen, Z.~Chen, Z.~Cui and L.~Dai, \textit{et al.}
Front. Phys. (Beijing) \textbf{16}, no.6, 64701 (2021)
doi:10.1007/s11467-021-1062-0
[arXiv:2102.09222 [nucl-ex]].

\bibitem{Bianconi:2015owa}
A.~Bianconi and E.~Tomasi-Gustafsson,
Phys. Rev. Lett. \textbf{114}, no.23, 232301 (2015)
doi:10.1103/PhysRevLett.114.232301
[arXiv:1503.02140 [nucl-th]].

\bibitem{Bianconi:2015vva}
A.~Bianconi and E.~Tomasi-Gustafsson,
Phys. Rev. C \textbf{93}, no.3, 035201 (2016)
doi:10.1103/PhysRevC.93.035201
[arXiv:1510.06338 [nucl-th]].

\bibitem{Lorenz:2015pba}
I.~T.~Lorenz, H.~W.~Hammer and U.~G.~Mei\ss{}ner,
Phys. Rev. D \textbf{92}, no.3, 034018 (2015)
doi:10.1103/PhysRevD.92.034018
[arXiv:1506.02282 [hep-ph]].

\bibitem{Belushkin:2006qa}
M.~A.~Belushkin, H.~W.~Hammer and U.~G.~Meissner,
Phys. Rev. C \textbf{75}, 035202 (2007)
doi:10.1103/PhysRevC.75.035202
[arXiv:hep-ph/0608337 [hep-ph]].

\bibitem{Lin:2021umk}
Y.~H.~Lin, H.~W.~Hammer and U.~G.~Mei\ss{}ner,
Phys. Lett. B \textbf{816}, 136254 (2021)
doi:10.1016/j.physletb.2021.136254
[arXiv:2102.11642 [hep-ph]].

\bibitem{Lin:2021umz}
Y.~H.~Lin, H.~W.~Hammer and U.~G.~Mei\ss{}ner,
Eur. Phys. J. A \textbf{57}, no.8, 255 (2021)
doi:10.1140/epja/s10050-021-00562-0
[arXiv:2106.06357 [hep-ph]].

\bibitem{Lin:2021xrc}
Y.~H.~Lin, H.~W.~Hammer and U.~G.~Mei\ss{}ner,
Phys. Rev. Lett. \textbf{128}, no.5, 052002 (2022)
doi:10.1103/PhysRevLett.128.052002
[arXiv:2109.12961 [hep-ph]].

\bibitem{Iachello:1972nu}
F.~Iachello, A.~D.~Jackson and A.~Lande,
Phys. Lett. B \textbf{43}, 191-196 (1973)
doi:10.1016/0370-2693(73)90266-9

\bibitem{BaBar:2013ukx}
J.~P.~Lees \textit{et al.} [BaBar],
Phys. Rev. D \textbf{88}, no.7, 072009 (2013)
doi:10.1103/PhysRevD.88.072009
[arXiv:1308.1795 [hep-ex]].

\bibitem{BaBar:2005pon}
B.~Aubert \textit{et al.} [BaBar],
Phys. Rev. D \textbf{73}, 012005 (2006)
doi:10.1103/PhysRevD.73.012005
[arXiv:hep-ex/0512023 [hep-ex]].

\bibitem{BESIII:2021dfy}
M.~Ablikim \textit{et al.} [BESIII],
[arXiv:2103.12486 [hep-ex]].

\bibitem{Tomasi-Gustafsson:2020vae}
E.~Tomasi-Gustafsson, A.~Bianconi and S.~Pacetti,
Phys. Rev. C \textbf{103}, no.3, 035203 (2021)
doi:10.1103/PhysRevC.103.035203
[arXiv:2012.14656 [hep-ph]].

\bibitem{Ellis:2001xc}
J.~R.~Ellis and M.~Karliner,
New J. Phys. \textbf{4}, 18 (2002)
doi:10.1088/1367-2630/4/1/318
[arXiv:hep-ph/0108259 [hep-ph]].

\bibitem{Tomasi-Gustafsson:2005svz}
E.~Tomasi-Gustafsson, F.~Lacroix, C.~Duterte and G.~I.~Gakh,
Eur. Phys. J. A \textbf{24}, 419-430 (2005)
doi:10.1140/epja/i2005-10030-6
[arXiv:nucl-th/0503001 [nucl-th]].

\bibitem{Haidenbauer:2014kja}
J.~Haidenbauer, X.~W.~Kang and U.~G.~Mei\ss{}ner,
Nucl. Phys. A \textbf{929}, 102-118 (2014)
doi:10.1016/j.nuclphysa.2014.06.007
[arXiv:1405.1628 [nucl-th]].

\bibitem{BaBar:2013ves}
J.~P.~Lees \textit{et al.} [BaBar],
Phys. Rev. D \textbf{87}, no.9, 092005 (2013)
doi:10.1103/PhysRevD.87.092005
[arXiv:1302.0055 [hep-ex]].

\bibitem{BESIII:2019hdp}
M.~Ablikim \textit{et al.} [BESIII],
Phys. Rev. Lett. \textbf{124}, no.4, 042001 (2020)
doi:10.1103/PhysRevLett.124.042001
[arXiv:1905.09001 [hep-ex]].

\bibitem{BESIII:2021aer}
M.~Ablikim \textit{et al.} [BESIII],
Phys. Lett. B \textbf{820}, 136557 (2021)
doi:10.1016/j.physletb.2021.136557
[arXiv:2105.14657 [hep-ex]].

\bibitem{Zyla:2020zbs}
P.~A.~Zyla \textit{et al.} [Particle Data Group],
PTEP \textbf{2020}, no.8, 083C01 (2020)
doi:10.1093/ptep/ptaa104

\bibitem{deMelo:2008rj}
J.~P.~B.~C.~de Melo, T.~Frederico, E.~Pace, S.~Pisano and G.~Salme,
Phys. Lett. B \textbf{671}, 153-157 (2009)
doi:10.1016/j.physletb.2008.11.056
[arXiv:0804.1511 [hep-ph]].

\bibitem{deMelo:2005cy}
J.~P.~B.~C.~de Melo, T.~Frederico, E.~Pace and G.~Salme,
Phys. Rev. D \textbf{73}, 074013 (2006)
doi:10.1103/PhysRevD.73.074013
[arXiv:hep-ph/0508001 [hep-ph]].

\bibitem{Milstein:2018orb}
A.~I.~Milstein and S.~G.~Salnikov,
Nucl. Phys. A \textbf{977}, 60-68 (2018)
doi:10.1016/j.nuclphysa.2018.06.002
[arXiv:1804.01283 [hep-ph]].

\bibitem{BESIII:2020vtu}
M.~Ablikim \textit{et al.} [BESIII],
Phys. Rev. Lett. \textbf{124}, no.11, 112001 (2020)
doi:10.1103/PhysRevLett.124.112001
[arXiv:2001.04131 [hep-ex]].

\bibitem{Bugg:2004rj}
D.~V.~Bugg,
Eur. Phys. J. C \textbf{36}, 161-168 (2004)
doi:10.1140/epjc/s2004-01955-5
[arXiv:hep-ph/0406292 [hep-ph]].

\bibitem{Cao:2018kos}
X.~Cao, J.~P.~Dai and Y.~P.~Xie,
Phys. Rev. D \textbf{98}, no.9, 094006 (2018)
doi:10.1103/PhysRevD.98.094006
[arXiv:1808.06382 [hep-ph]].

\bibitem{Wang:2021abg}
L.~M.~Wang, S.~Q.~Luo and X.~Liu,
Phys. Rev. D \textbf{105}, no.3, 034011 (2022)
doi:10.1103/PhysRevD.105.034011
[arXiv:2109.06617 [hep-ph]].

\bibitem{Sakharov:1948plh}
A.~D.~Sakharov,
Zh. Eksp. Teor. Fiz. \textbf{18}, 631-635 (1948)
doi:10.1070/PU1991v034n05ABEH002492

\bibitem{Farrar:1975yb}
G.~R.~Farrar and D.~R.~Jackson,
Phys. Rev. Lett. \textbf{35}, 1416 (1975)
doi:10.1103/PhysRevLett.35.1416

\bibitem{Chernyak:1984bm}
V.~L.~Chernyak and I.~R.~Zhitnitsky,
Nucl. Phys. B \textbf{246}, 52-74 (1984)
doi:10.1016/0550-3213(84)90114-7

\bibitem{Alexandrou:2018sjm}
C.~Alexandrou, S.~Bacchio, M.~Constantinou, J.~Finkenrath, K.~Hadjiyiannakou, K.~Jansen, G.~Koutsou and A.~Vaquero Aviles-Casco,
Phys. Rev. D \textbf{100}, no.1, 014509 (2019)
doi:10.1103/PhysRevD.100.014509
[arXiv:1812.10311 [hep-lat]].

\bibitem{Kim:1995mr}
H.~C.~Kim, A.~Blotz, M.~V.~Polyakov and K.~Goeke,
Phys. Rev. D \textbf{53}, 4013-4029 (1996)
doi:10.1103/PhysRevD.53.4013
[arXiv:hep-ph/9504363 [hep-ph]].

\bibitem{BESIII:2020uqk}
M.~Ablikim \textit{et al.} [BESIII],
Phys. Lett. B \textbf{814}, 136110 (2021)
doi:10.1016/j.physletb.2021.136110
[arXiv:2009.01404 [hep-ex]].

\bibitem{BaBar:2007fsu}
B.~Aubert \textit{et al.} [BaBar],
Phys. Rev. D \textbf{76}, 092006 (2007)
doi:10.1103/PhysRevD.76.092006
[arXiv:0709.1988 [hep-ex]].

\bibitem{BESIII:2021ccp}
M.~Ablikim \textit{et al.} [BESIII],
Phys. Rev. D \textbf{104}, no.9, L091104 (2021)
doi:10.1103/PhysRevD.104.L091104
[arXiv:2108.02410 [hep-ex]].

\bibitem{Belitsky:2001ns}
A.~V.~Belitsky, D.~Mueller and A.~Kirchner,
Nucl. Phys. B \textbf{629}, 323-392 (2002)
doi:10.1016/S0550-3213(02)00144-X
[arXiv:hep-ph/0112108 [hep-ph]].

\bibitem{ParticleDataGroup:2020ssz}
P.~A.~Zyla \textit{et al.} [Particle Data Group],
PTEP \textbf{2020}, no.8, 083C01 (2020)
doi:10.1093/ptep/ptaa104

\bibitem{BESIII:2014fwz}
M.~Ablikim \textit{et al.} [BESIII],
Phys. Lett. B \textbf{735}, 101-107 (2014)
doi:10.1016/j.physletb.2014.06.013
[arXiv:1403.6011 [hep-ex]].


\end{thebibliography}
\end{document}